\documentclass{article}
\usepackage{spconf,amsmath,graphicx}
\usepackage{spconf,amsmath,graphicx}
\usepackage{multirow}
\usepackage{float}
\usepackage{here}
\usepackage{pgfplots}
\usepackage{latexsym}
\usepackage{url}
\usepackage{amsthm}
\usepackage{amssymb}
\usepackage{graphicx}
\usepackage{booktabs}
\usepackage{placeins}
\usepackage{tabularx}
\usepackage{epstopdf}
\usepackage{color}
\usepackage{subcaption}
\usepackage{amsfonts}
\usepackage{amsmath}
\usepackage{balance}
\usepackage{enumitem}
\usepackage{xcolor}
\usepackage{wrapfig}
\usepackage{graphics}
\usepackage{todonotes}
\usepackage{makecell}
\usepackage{bbm}
\usepackage{bm}

\title{MASR: Multi-label Aware Speech Representation}
\name{Anjali Raj$^1$, Shikhar Bharadwaj$^1$, Sriram Ganapathy$^{1,2}$, Min Ma$^1$, Shikhar Vashishth$^1$}
\address{\small{\texttt{\{rajanjali,shikharop,srigana,minm,shikharv\}@google.com}}\\
$^1$Google Research, $^2$Indian Institute of Science}
\begin{document}
\ninept
\maketitle

\newcommand{\refalg}[1]{Algorithm \ref{#1}}
\newcommand{\refeqn}[1]{Equation \ref{#1}}
\newcommand{\reffig}[1]{Figure \ref{#1}}
\newcommand{\reftbl}[1]{Table \ref{#1}}
\newcommand{\refsec}[1]{Section \ref{#1}}

\newcommand{\reminder}[1]{\textcolor{red}{[[ #1 ]]}\typeout{#1}}
\newcommand{\reminderR}[1]{\textcolor{gray}{[[ #1 ]]}\typeout{#1}}

\newcommand{\add}[1]{\textcolor{red}{#1}\typeout{#1}}
\newcommand{\remove}[1]{\sout{#1}\typeout{#1}}

\newcommand{\method}{MASR}
\newcommand{\systemfull}{Metadata Aware Speech Representation Learning}
\newcommand{\system}{MASR}

\newcommand{\mc}[1]{\mathcal{#1}}
\newcommand{\bmm}[1]{\bm{\mathcal{#1}}}
\newcommand{\real}[1]{\mathbb{R}^{#1}}

\newcommand{\tensor}{\mathcal{X}}
\newcommand{\Real}{\mathbb{R}}

\newcommand{\tuples}{\mathbb{T}}

\newcommand\norm[1]{\left\lVert#1\right\rVert}

\newcommand{\note}[1]{\textcolor{blue}{#1}}
\newcommand{\review}[1]{{#1}}

\newcommand*{\Scale}[2][4]{\scalebox{#1}{$#2$}}%
\newcommand*{\Resize}[2]{\resizebox{#1}{!}{$#2$}}%
\def\mat#1{\mbox{\bf #1}}
\begin{abstract}
In the recent years, speech representation learning is constructed primarily as a self-supervised learning (SSL) task, using the raw audio signal alone, while ignoring the side-information that is often available for a given speech recording. 
In this paper, we propose \textbf{\method{}}, a \textbf{M}ulti-label \textbf{A}ware \textbf{S}peech \textbf{R}epresentation learning framework, which addresses the aforementioned limitations.
\method{} enables the inclusion of multiple external knowledge sources to enhance the utilization of meta-data information. The external knowledge sources are incorporated in the form of sample-level pair-wise similarity matrices that are useful in a hard-mining loss. 
A key advantage of the \system{} framework is that it can be combined with any choice of SSL method.
Using \method{} representations, we perform evaluations on several downstream tasks such as language identification,  speech recognition and other non-semantic tasks such as speaker and emotion recognition. In these experiments, we illustrate significant performance improvements for the \system{} over other established benchmarks.  We perform a detailed analysis on the language identification task to provide insights on how the proposed loss function enables the representations to separate closely related languages. 
\end{abstract}

\section{Introduction}
\label{sec:intro}
In recent years,  representation learning of speech signals has seen a paradigm shift from knowledge-driven features such as mel-frequency cepstral coefficients (MFCC) \cite{davis1980comparison} to data driven  learning approaches like wav2vec \cite{wave2vec}.   Representation learning of audio signals, from the raw data, has closely mirrored the advances made in self-supervised learning (SSL) of other domains such as natural language processing (NLP) \cite{bert} and computer vision \cite{vision_unsup1}. 
The initial approaches for SSL of speech data explored the modeling of the continuous latent representations of speech.   The popular examples include constrastive predictive coding (CPC)  \cite{oord2018representation}, wav2vec modeling \cite{wave2vec} and auto-regressive predictive coding (APC)   \cite{chung2020generative}. In the subsequent modeling frameworks, a discretized representation of audio is found to be more beneficial  than the continuous space \cite{nguyen2022discrete}. The recent success in SSL speech modeling  relies on three major steps - i) converting speech to a discrete token sequence, ii) masking portions of the speech signal and, iii) predicting the token sequence corresponding to the masked parts of the audio. Examples of such approaches include wav2vec 2.0 \cite{wave2vec2}, hidden unit bidirectional encoder representation from transformers (HuBERT) \cite{hubert} and best random quantizer (BestRQ) \cite{best_rq}. 
The downstream tasks for these models were initially centered on low resource automatic speech recognition (ASR) \cite{schneider2019wav2vec} and multilingual ASR \cite{best_rq}. 
However, these approaches have been extended to various other downstream tasks such as emotion recognition \cite{emotion_recog}, speaker recognition \cite{speaker_recog}, language identification \cite{langid} and other audio tasks through the SUPERB \cite{yang2021superb} and NOSS \cite{trill} challenges. 

\begin{figure*}[!t]
\center
\includegraphics[scale=0.9]{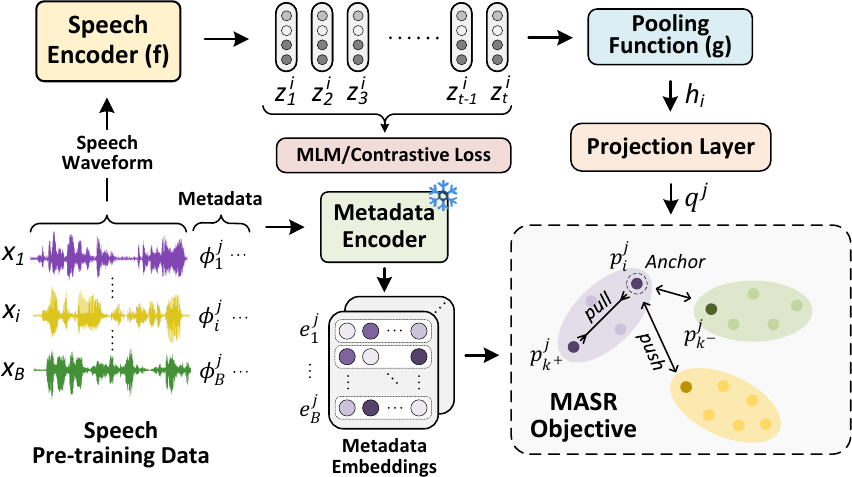}
\caption{\label{fig:overview}\small Overview of \method{} framework. For a speech sample ($x_i$) in a batch of speech samples, \method{} utilizes a speech encoder ($f$) to obtain frame-level representations ${z^i_1, z^i_2, ..., z^i_t}$. These are fed to a pooling function ($g$) to derive utterance-level embedding $h_i$. For the jth metadata, the metadata label $\phi_i^j$ is encoded to real-valued vectors, $e_i^j$, while the utterance-embedding $h_i$ is transformed to $q_i^j$ through a projection layer. The  \method{} loss is then computed on the transformed representations.}
	\vspace{-0.15in}
\end{figure*}

Typically, speech data resources possess supplementary meta-data alongside the audio signals, manifesting in various forms such as speaker, language, emotion, style, and transcripts. Often, during speech data collection or data mining from the web, additional information is recorded along the semantic and non-semantic aspects of the data. The current representation learning paradigms completely disregard these additional streams of information. 

The prior works \cite{talnikar2021joint,bai2022joint} have explored the joint optimization of SSL loss function with speech recognition loss functions. Further, the previous work by Vashisth et al. \cite{lasr} investigated an approach for combining SSL loss functions with a language label based triplet mining loss. Nevertheless, previous studies have not investigated the potential of employing multiple meta-labels for the purpose of representation learning.   In this paper, we propose \method{}, a \underline{\textbf{M}}ulti-label \underline{\textbf{A}}ware \underline{\textbf{S}}peech \textbf{R}epresentation learning framework, which attempts to address some of these limitations by leveraging the existence of multiple meta-information sources and external knowledge sources.
Our contributions in this work can be summarized as follows.

\begin{itemize}[itemsep=2pt,parsep=2pt,partopsep=2pt,leftmargin=*,topsep=2pt]
    \item We propose \method{}, a novel framework for incorporating multiple meta-labels in any self-supervised learning (SSL) framework.
    \item \method{} enables inclusion of external knowledge sources to enhance the utilization of meta-data information.
    \item Through extensive experiments on several downstream tasks, we demonstrate the effectiveness of our proposed approach. 
\end{itemize}

\section{Related Works}
\label{sec:relatedwork}

\textbf{Speech Self-Supervised Learning:} The initial works in speech representation learning used  a series of convolutional and long-short term memory (LSTM)  layers \cite{oord2018representation}  with a contrastive loss that  predicts the future frames of the audio signal \cite{chung2020generative}. The first in the series of wav2vec models \cite{wave2vec} started with similar contrastive prediction losses, while the  subsequent version \cite{wave2vec_c}  also involved a discretization of the audio using a vector quantization module. 
The recent version in this series, wav2vec 2.0 \cite{wave2vec2} also included  transformer based encoding layers. The application of MLM loss with wav2vec model, called w2v-bert \cite{chung2021w2v}, combined the discretization of speech with the masked modeling. Another notable effort, titled HuBERT \cite{hubert}, performed iterative clustering/discretization and masked language modeling. A simplification of  k-means clustering algorithm on the hidden layer representations, allowed the direct encoding of the input mel-spectrogram with a random quantizer \cite{best_rq}. All these methods perform frame-level encoding with the benefits illustrated primarily for   speech  recognition tasks  \cite{baevski2021unsupervised} or zero-resource language modeling task \cite{maekaku2022exploration}.  The work proposed in this paper  investigates the combination of utterance level objective functions   with frame-level MLM loss functions. 

\noindent \textbf{Speech Supervised Learning:} In the direction of utterance level embedding extractors, the   supervised embedding extraction using neural networks were explored for speaker recognition \cite{heigold2016end} and language recognition \cite{snyder2018spoken}. 
The  architectures explored for this supervised setting include  time-delay neural network (TDNN)  \cite{snyder2018x}, residual networks \cite{li2017deep} and TDNN with enhanced channel attention (ECAPA-TDNN)  \cite{desplanques2020ecapa}.  In contrast to the supervised embedding extraction frameworks, this work explores the combination of self-supervision and meta-data based hard-mining losses. 

\noindent \textbf{Non-semantic Speech Representation:} 
TRILL \cite{trill} leverages temporal proximity as a supervision signal for learning non-semantic representation. The TRILL and subsequent variants such as FRILL \cite{frill} and TRILLsson \cite{trillson} have shown encouraging results on NOSS (non-semantic speech) tasks \cite{trill}.  The effort discussed in COLA \cite{cola}  changes the negative sampling strategy for audio tasks. While these works are designed for contrastive learning, our work explores a combination of meta-data based weak supervision losses that can be coupled with any SSL   approach.  

 

\noindent \textbf{Joint Learning:} For joint modeling of self-supervised and supervised objectives in speech processing, the combined loss of CTC (connectionist temporal classification) and CPC losses for ASR have been explored \cite{talnikar2021joint}. 
The work referred to as UniSpeech \cite{wang2021unispeech} combined CTC loss with a phonetic loss, while JOIST (joint learning of supervised and unsupervised losses) \cite{bai2022joint} used the mixture of MLM and speech recognition loss functions. 
A recent work exploring the meta-data information of language, called label aware speech representation learning (LASR) \cite{lasr}, showed the benefits of combining meta-data of language labels with the SSL framework. 
However, these methods are restricted to a particular type of meta-information, while this work proposes a general framework which enables utilization of multiple streams of meta-data.

\section{Proposed Approach}
\label{sec:methodology}
A detailed illustration of the complete framework  is provided in Figure \ref{fig:overview}. Our framework is a multi-label extension of the LASR \cite{lasr} method, that enabled the inclusion of language meta-information during self-supervised learning. The LASR approach is not directly generalizable to the presence of multiple meta-information streams. 
Further, the LASR setting does not allow the incorporation of meta-information in the form of soft-labels.
\method{} addresses these limitations by introducing two additional modules in the LASR framework: (i) Metadata Encoder, which allows inclusion of external knowledge sources, and  (ii) Projection Layer, which allows the utilization of multiple  meta-information streams jointly. 

\subsection{Notations}
We denote a speech pre-training dataset with meta-information as $\mc{D} = \{(\bmm{X}_1,  \Phi_1), (\bmm{X}_2,  \Phi_2), ..., (\bmm{X}_N,  \Phi_N)\}$ where $\bmm{X}_i$ denotes the waveform and $\Phi_i = \{\phi_i^1, \phi_i^2, ..., \phi_i^M\}$ represents the meta-information associated with the $i$-th sample. A self-supervised speech-representation learning model, such as w2v-BERT \cite{wave2vec}  \cite{w2v_bert} or BEST-RQ \cite{best_rq}, is defined as an encoder, $f: \bmm{X} \rightarrow \bmm{Z}$, which transforms a waveform $\bmm{X}$ to its frame-level representation $\bmm{Z} = [\bm{z}_1, \bm{z}_2, ..., \bm{z}_T]$. We utilize an aggregation model $g: \bmm{Z} \rightarrow \bm{h}$ to obtain an utterance level embedding. In this work, we define $g$ as an average pooling operation, i.e., $\bm{h} = g(\bmm{Z}) = \dfrac{1}{T} \sum_{t}{\bm{z}_t}.$

\subsection{Metadata Encoder (ME)}
Most of the existing methods utilize metadata information directly in their raw form.
In this work, we introduce a metadata encoder module that leverages external knowledge resources to generate a representation for each type of  metadata. Specifically, 
each type of meta-data is mapped to a real-valued $d$-dimensional space. Let the mapping be $\bm{\pi_j}: \boldsymbol {\phi} ^j \rightarrow \bm {e} ^j $, where $j$ denotes the metadata type, and $\bm{e}^j$ denotes the encoding of metadata.  
Thus, for each metadata, the encoder module gives a metadata specific representation $\bm{e}^j$, which is utilized for computing the final objective function. In this work, metadata encoder $\bm{\pi _j}$ is pre-defined and not learnable.

\subsection{Projection Layer}
In order to integrate multiple types of metadata information into MASR, we introduce a metadata specific projection applied to the utterance-level representation $\bm{h}$. 
We note that using $\bm{h}$ directly, makes the overall speech encoder   specific to a particular type of metadata. Moreover, it may lead to conflicting objectives when using multiple types of metadata. To overcome this, we introduce a metadata-specific transformation function, $\bm{\delta_j}: \bm{h} \rightarrow \bm {q}^j$, where $j$ denotes the metadata type, and $\bm {q}^j \in \mathbb {R}^d$ denotes the projection. Finally, the metadata-specific aggregated representation $\bm{\mathrm {p}}^j$ is obtained by concatenating the metadata encoder and the projection layer outputs, i.e.,
\begin{equation}\label{eq:h_new}
\bm{\mathrm {p}}^j = [ \bm {q}^j; \alpha_j \bm {e} ^j ], \ \  \forall j \in [1 ... M]
\end{equation} 
where $\alpha_j$ is used to scale the importance of $\bm{e}^j$ with respect to $\bm{q}^j$. Here, $M$ denotes the number of  meta-data streams. 
In this study, we employ $\bm{\delta_j}$ as a fully connected layer for each metadata.

\subsection{Training Loss}
In a batch of size $B$, for   $i^{th}$ speech utterance and $j^{th}$ metadata type, we define a positive and negative sample set $\rho^+_{ij}$ and $\rho^-_{ij}$ such that,
\begin{eqnarray}
\rho^+_{ij} &=& \{k \ | \ \phi^j_i = \phi^j_k, \ \  \forall k \in [0, ...,  B-1]\}, \nonumber \\ 
\rho^-_{ij} &=& \{k \ | \ \phi^j_i \neq \phi^j_k, \ \  \forall k \in [0, ...,  B-1] \} \nonumber    
\end{eqnarray}
We utilize the hard triplet mining loss  \cite{loss_hard_triplet,triplet_loss_lang}, where the farthest (cosine distance) positive sample (from $\rho^+_{ij}$) and the nearest negative sample (from $\rho^-_{ij}$) are used for computing the triplet loss, defined as,


\begin{equation}\label{eq:hard}
\small
\mc{L}_{\texttt{META}}^j = \sum_{i}  [\gamma + d(\bm{q}^j_{i}, \bm{q}^j_{k+}) - d (\bm{q}^j_{i}, \bm{q}^j_{k-})]_+
\end{equation} 
where,
\begin{eqnarray}\label{eqn:sample_sel}
k^+ = \textrm{argmax}_{k \in \rho^+_{ij} } d(\bm{\mathrm {p}}^j_i, \bm{\mathrm {p}}^j_k) ~~;~~
k^- = \textrm{argmin} _{k \in \rho^-_{ij} } d(\bm{\mathrm {p}}^j_i, \bm{\mathrm {p}}^j_k)
\end{eqnarray}
with $k^+$ denoting the farthest positive sample (based on representations, $\bm{\mathrm {p}}^j$) and $k^-$ denotes the nearest negative sample. The final loss function is given by, 
\begin{equation} \label{eq:hard-triplet}
\mc{L} _{\texttt{\method{}}} = \mc{L}_{\texttt{SSL}} + \sum _j \lambda _j \cdot \mc{L}_{\texttt{META}}^j.
\end{equation}
Here, $\mc{L}_{\texttt{SSL}}$ is the loss corresponding to the self-supervised speech encoding method $f$, and $\lambda _j$ decides the trade-off between the SSL objective and hard-triplet objective for the $j$-th meta-data stream.

{In our work, we have primarily explored meta-data  of language labels encoded in various ways defined in URIEL \cite{littell2017uriel} database}. Further, we also perform ablation studies on NOSS evaluations using text transcripts and speaker's geographic location information.

\begin{table*}[t!]
	\centering
	\begin{tabular}{lcccccccccc}
	    \toprule
	     & \multicolumn{5}{c}{\textbf{FLEURS}} & \multicolumn{5}{c}{\textbf{Dhwani}}  \\
	    \cmidrule(r){2-6} \cmidrule(r){7-11}
	    \multicolumn{1}{l}{\textbf{Method}} & O (48) & NO (54) & \multicolumn{3}{c}{Overall}  & O (5) & NO (17) & \multicolumn{3}{c}{Overall} \\
	    \cmidrule(r){2-2} \cmidrule(r){3-3} \cmidrule(r){4-6} \cmidrule(r){7-7} \cmidrule(r){8-8} \cmidrule(r){9-11}
	     & Acc & Acc & Acc & F1 & EER$\downarrow$ & Acc  & Acc & Acc & F1 & EER$\downarrow$ \\
		\midrule
		
		w2v-BERT \cite{chung2021w2v} & 87.7 & 69.6 & 78.0 & 77.7 & 0.5 & 78.8 & 49.9 & 58.0 & 42.6 & 15.4 \\
        \ \ + LASR \cite{lasr} & 88.9 & 74.3 & 81.3 & 80.4 & 0.5 & 78.1 & 52.2 & 59.5 & 44.2 & \textbf{15.2} \\
        \ \ + \method{} & \textbf{91.4}&\textbf{76.3}& \textbf{83.4}& \textbf{81.3}&\textbf{0.4}&\textbf{81.0}&\textbf{52.4}& \textbf{60.7}&\textbf{46.2}&15.3 \\
        \midrule  
        BEST-RQ \cite{best_rq} &  85.6 & 65.2 & 75.4 & 72.8 & 0.9 & 76.2 & 46.4 & 54.7 & 39.8 & 16.9 \\
		\ \ + LASR \cite{lasr} & 90.6 & 73.4 & 81.6 & 79.7 & 0.5 & 77.0 & 48.6 & 57.7 & 43.0 & 16.1 \\
		
		\ \ + \method{} & \textbf{91.3} & \textbf{74.6} & \textbf{83.7} & \textbf{81.5} & \textbf{0.3} & \textbf{80.6} &	\textbf{50.6}	 &  \textbf{59.6} &\textbf{44.2} & \textbf{15.8}		 \\
		\bottomrule 
	\end{tabular}
	\caption{\label{tbl:langid_main} Language identification accuracy (\%), macro-F1 and equal error rate (EER) for various approaches. \textit{O} stands for languages that overlap with the pre-training data and \textit{NO} are the non-overlapping languages. In parenthesis, we report the number of classes in each category. The downward arrow for a label indicates lower values are better. 
	Refer to Section \ref{sec:results} for details.}
\end{table*}

\section{Experimental Setup}
\label{sec:expt}

\subsection{Pre-training Data} 

In our experimental setup, we utilize a substantial collection of publicly available speech data for pre-training, amounting to approximately 429k hours of audio. This encompasses various datasets, including 372k hours of speech data spanning 23 languages from the VoxPopuli \cite{dataset_vp}, 50k hours of speech in 25 languages from the Common Voice \cite{dataset_cv}, 50k hours of read speech in 8 European languages from the Multilingual LibriSpeech (MLS) \cite{dataset_mls}, and 1k hours of telephone conversation data encompassing 17 African and Asian languages from the BABEL dataset \cite{dataset_babel}. In total, our combined dataset comprises speech utterances from 75 different languages. The language metadata is available for the entire pre-training data.

\subsection{Evaluation Data}

\noindent \textbf{Language Identification}
We report language identification performance on two datasets - FLEURS \cite{dataset_fleurs} and Dhwani \cite{dataset_in_w2v}.
FLEURS is a n-way parallel speech dataset with uniform coverage of $102$ languages. 
On average, each language contains about $12$ hours of audio from multiple speakers.
The dataset contains sentences sourced from Wikipedia \cite{dataset_flores}.
For fine-tuning the pre-trained models, we use the standard FLEURS train set and report performance on the test set. In order to understand the robustness of the model for language recognition, we also perform experiments using the Dhwani dataset \cite{dataset_in_w2v}.
The Dhwani dataset contains raw audio data acquired from social media, video-sharing and news platforms like YouTube and News On Air bulletins.
This corpus has around $12.6$k hours of publicly downloadable speech data from $22$ Indian languages. \\

\noindent \textbf{Automatic Speech Recognition:}
For the ASR task, we have utilized the Multilingual LibriSpeech (MLS) corpus, which comprises of $50$k hours of read speech from LibriVox corpus across $8$ European languages.
We use the standard train (full) for fine-tuning the models, and report the results on the test splits of MLS. \\ 

\noindent \textbf{Non-semantic Speech Benchmark:}
The third set of evaluation uses $6$ tasks defined as part of the NOSS benchmark \cite{trill}. This captures emotion recognition, speaker recognition, language identification, environment sound classification and masked speech classification.
Unlike the LangID and ASR tasks defined above, the pre-trained models are not fine-tuned for the downstream tasks in the NOSS benchmark.
A randomly initialized soft-max layer is alone trained for each downstream task in the NOSS suite.

\subsection{Implementation Details}
For most of our ablation experiments, we use the SSL approach of BEST-RQ \cite{best_rq} as the baseline choice.
The baseline system is trained for $1.25$M steps on the pre-trained data.
The \system $~$models are initialized with the BEST-RQ model trained for $1$M steps, followed by $250$k steps of additional pre-training using the proposed objective functions.
The hyper-parameter $\lambda = 16$ has been adopted   for the language related metadata.


\subsection{Language meta-data encoder}
The encoding of the language labels is obtained from the lang2vec models extracted from the URIEL \cite{littell2017uriel} typological database. 
The dataset encompasses a range of metrics that capture the language features derived from the syntactic structures of the language, as well as features derived from the World Atlas of Language Structures \cite{wals}. 
The set comprises of a vast number of languages, totaling 8070. In our case, we use the lang2vec related to the $75$ pre-training languages.
The distances between languages is the cosine similarity between the respective feature vectors. 

\section{Results} 
\label{sec:results}

\begin{table*}[t!]
	\centering
	\begin{tabular}{lcccccccccc}
	    \toprule
	     \multicolumn{1}{l}{\textbf{Method}} & \multicolumn{1}{c}{\textbf{LangID}} & \multicolumn{1}{c}{\textbf{Speaker Verification}} &
	     \multicolumn{1}{c}{\textbf{Emotion Recogition}} &
	     \multicolumn{3}{c}{\textbf{Audio Classification}}\\
	    \cmidrule(r){2-2} \cmidrule(r){3-3} \cmidrule(r){4-4} \cmidrule(r){5-7}
	     & \textbf{VoxForge} & \textbf{ASVSpoof2019} & \textbf{Iemocap} & \textbf{Mask Challenge} & \textbf{Esc50-human} & \textbf{Esc50-cough} \\ 
		\midrule
		
        BEST-RQ \cite{best_rq} &  94.6  & 94.0 & 54.0 & 61.1  & 72.0 & 90.9 \\
        
		\ \ + LASR  & \textbf{96.0} & \textbf{97.9} & \textbf{60.7} & 58.1 
		& 63.6 & 87.9   \\
		\ \ + \method{} & 94.0 & 94.3  & 60.0 & \textbf{63.0} & 80.3 & \textbf{94.0}  \\
		\ \ + GeoMASR & 95.7  & 96.1 & 58.8  & 62.2 & \textbf{81.8} & 92.5  \\
		\ \ + TextMASR & 90.8 & 93.2 & 57.5 & 61.2  & \textbf{81.8} & 87.9  \\
		\bottomrule
	\end{tabular}
	\caption{\label{tbl:noss} Non Semantic Speech Tasks accuracy (\%) for various tasks. The table shows the average performance of the different embeddings on a number of downstream tasks with fixed conventional splits. We find that methods trained with the MASR objective achieves better performance in terms of accuracy in 3 out of 6 tasks when compared to BEST-RQ, and BEST-RQ + LASR. Refer to Section \ref{sec:results} for details.}
\end{table*}
\subsection{Language Identification Performance}
The language identification results are reported in Table \ref{tbl:langid_main}. 
The baselines used for comparison include  w2v-BERT \cite{w2v_bert} and BEST-RQ \cite{best_rq}.
The metrics include accuracy over the entire test set, along with the macro-F1 score and the EER (equal error rate). 
Using the \method{} framework, the performance is shown to improve on both the datasets, i.e., Dhwani and FLEURS.

To summarize, following are the observations from the results:
\begin{itemize}[itemsep=2pt,parsep=2pt,partopsep=2pt,leftmargin=*,topsep=2pt]
    \item The MASR approach improves the
BEST-RQ model relatively by $11.2$\%, $11.9$\%, and $66.6$\% in terms of accuracy, F1 and EER metrics, respectively. The corresponding improvements for the MASR framework over the LASR model are  by $2.6$\% , $2.3$\% and $40.0$\% in terms of accuracy, F1 and EER respectively. 
Similar improvement in performance is also observed for Dhwani dataset as well. The \method{} framework improves the relative accuracy over the BEST-RQ baseline by $9.0$\%, F1 by $11.1$\% and EER by $8.3$\%. 

\item This increase in performance is  also prominent in the SSL baseline with w2v-BERT framework. 
In addition, the overlapped (languages that are common between pre-training and fine-tuning) and the Non-overlapped (languages that are different from those in the pre-training dataset) class accuracy also show a similar trend across the two datasets.

\item The improvement in Non-Overlap accuracy is observed to be more for \method{} as compared to that in the Overlap accuracy, which reflects that the model is able to generalize the meta-data based objective functions to unseen languages.   

\item From our analysis, we find that \method{} influences the selection of negative samples by more than $80$\%, i.e., nearly $12$ out of $16$ samples in a batch undergo a change in their negative neighbour selection compared to the LASR training. 

\end{itemize}

\subsection{Evaluation on Non-Semantic Tasks}
The evaluation on NOSS benchmark was carried out on 6 tasks which include the Voxforge, Vox-filtered, Mask Challenge, Iemocap, ESC-human, ESC-cough  and ASVSpoof-2019. VoxForge is a language classification dataset that consists of user-submitted audio clips in six languages: English, Spanish, French, German, Russian, and Italian. ESC50-human is a benchmark dataset of 2k $5$-second environmental audio recordings, organized into $50$ semantic classes.

For each representation-task pair, we explore different downstream models, representation aggregation techniques, and normalization methods.
In \reftbl{tbl:noss}, we report results for various ablations of \method{} as well as the LASR method \cite{lasr}.
We observe that using LASR objective along with the SSL objective on BEST-RQ leads to a deterioration in performance on $3$ out of $6$ NOSS tasks.
Adding the metadata encoder with the projection layer leads to an increase in performance over the BEST-RQ model for $4$ out of $6$ tasks.
The description of the last two rows,  Geo\method{} and Text\method{}, is given in Section~\ref{sec:geomasr}.
The addition of Geological metadata as loss computation head massively improves the MASR performance on Voxforge which is a language identification task. It is interesting to observe that incorporating extra head reflecting Indic geography is able to improve Language identification task for European languages. Overall, MASR variants see more improvements in the  audio classification tasks.

\begin{figure}[!t]
    \includegraphics[width=0.9\columnwidth]{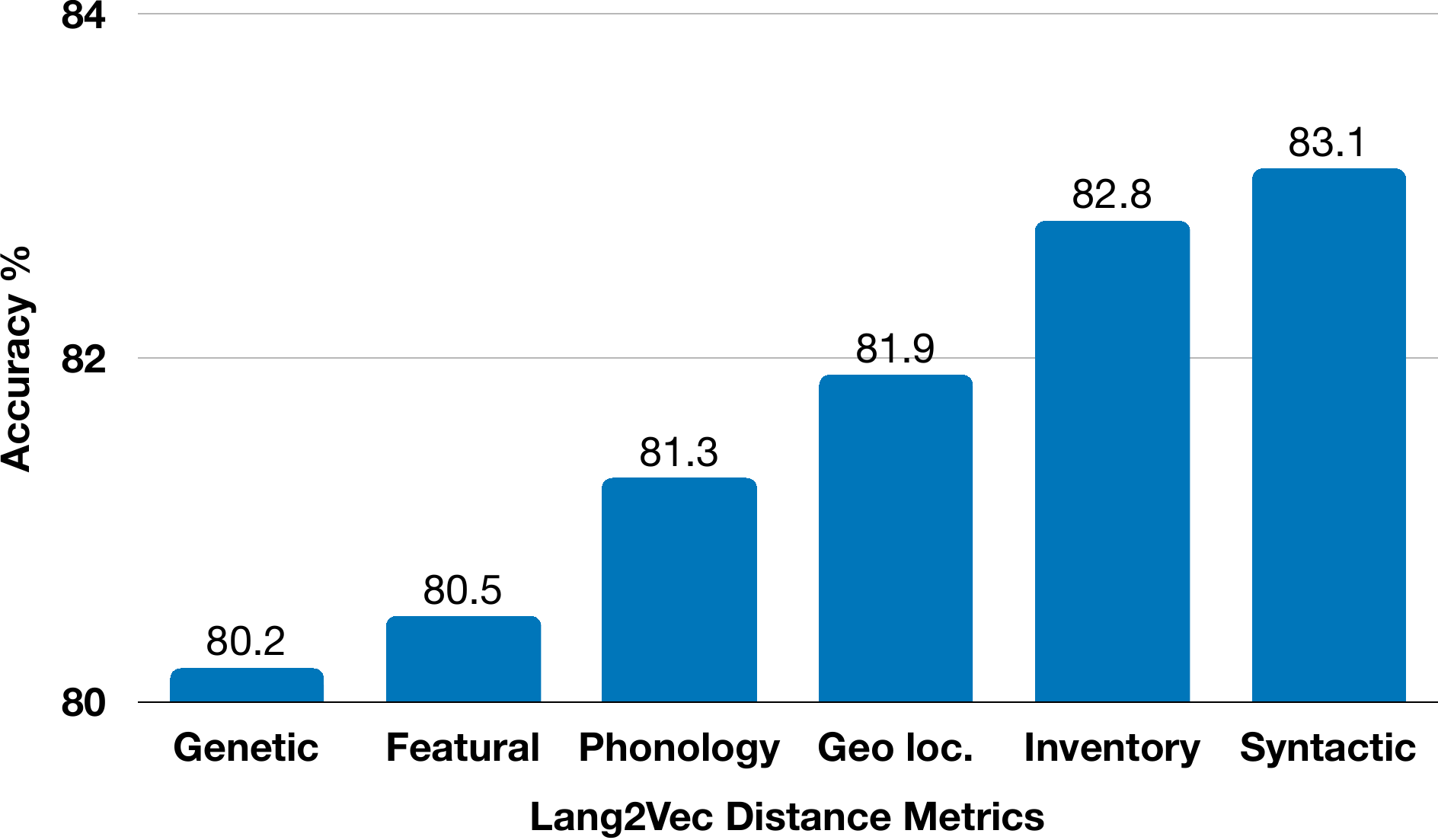}
    \caption{\label{fig:Lang2vec_dist} Comparison of different distance vectors from Lang2vec with combined and selection hard-triplet objective. For deciding the best metric, we used 250k checkpoint of the pre-trained models.}
\end{figure}


\subsection{Impact on Automatic Speech Recognition}
We evaluate the baseline system and the \method{} approach for the semantic ASR tasks defined on the MLS dataset. The word error rate (WER) results are reported in \reftbl{tbl:asr_main}.
All models are fine-tuned for $10$k steps. We observe that the MASR setting, even with the use of non-semantic language based meta-data, does not cause a degradation on the average WER.

\section{Analysis of Language Identification}
\label{sec:analysis}
\subsection{Choice of Meta-data encoder}

For \method{}, we explored a variety of ways to encode the language meta-data using various Lang2vec representations \cite{littell2017uriel}. In particular, we experiment with \textit{syntactic}, \textit{geographic}, \textit{phonetic}, \textit{featural}, \textit{genetic} and \textit{inventory} based embeddings as defined by  Littell et al. (2017). 
As shown in Figure \ref{fig:Lang2vec_dist}, \method{} relies significantly on the choice of  Lang2vec embeddings, with the syntactic encoding providing the best accuracy.
Thus, we use the syntactic distance for all the other  experiments in this paper.


\begin{figure*}[!t]
\center
\includegraphics[width=0.8\textwidth]{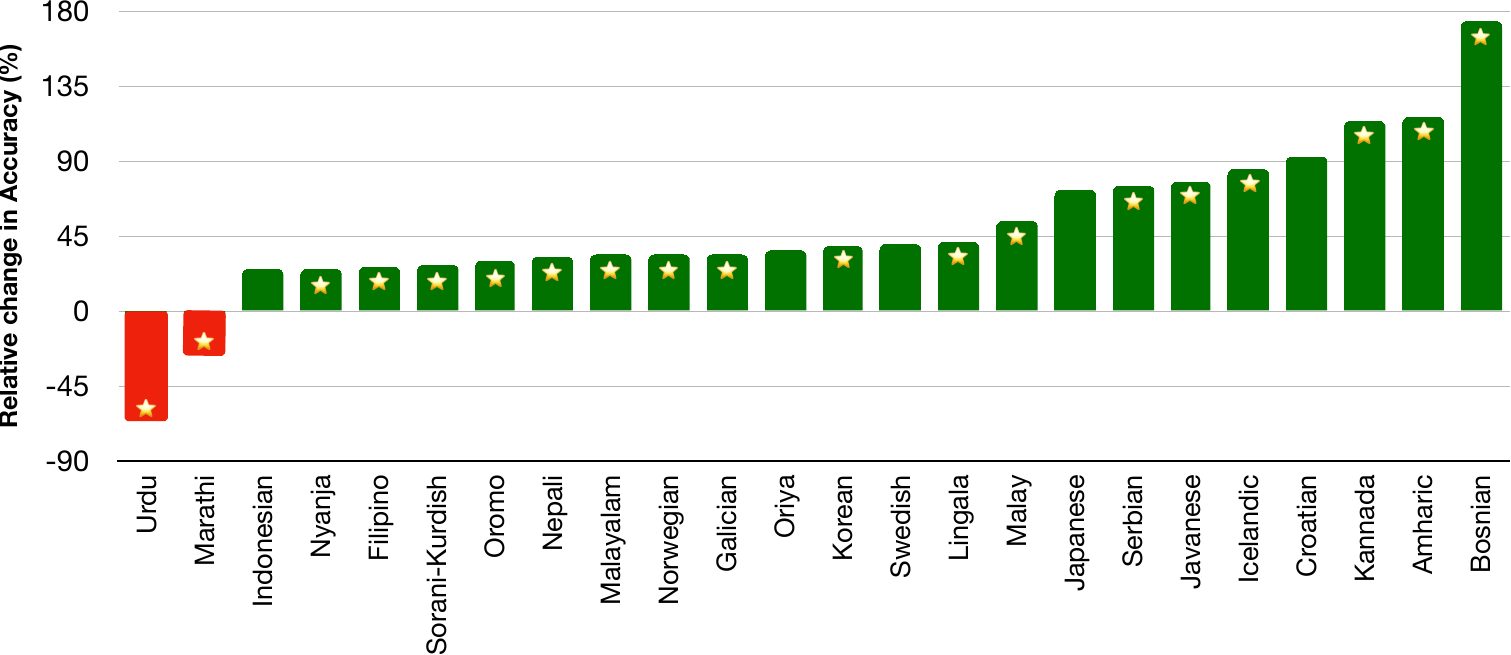}
\caption{\label{fig:delta}\small Relative changes in accuracy from the best MASR language identification model over the BEST-RQ baselines. Languages whose relative changes lie between $(-25\%, +25\%)$ have been skipped for easier readability. The star icon indicates non-overlapped languages.}
\end{figure*}

\subsection{Locale Level Improvements on FLEURS dataset}
Overall our single best langID model achieved \review{83.7\% accuracy over 102 FLEURS locales, which is 11\% better than the baseline BEST-RQ accuracy of 75.4\%}.
Locale wisely, 88 locales have been improved, while 13 locales observed regressions in accuracy. 
Figure~\ref{fig:delta} illustrates the details the significant changes in prediction for the \method{} over the BEST-RQ baseline.



\begin{figure}[t]
\center 
    \includegraphics[width=0.9\columnwidth]{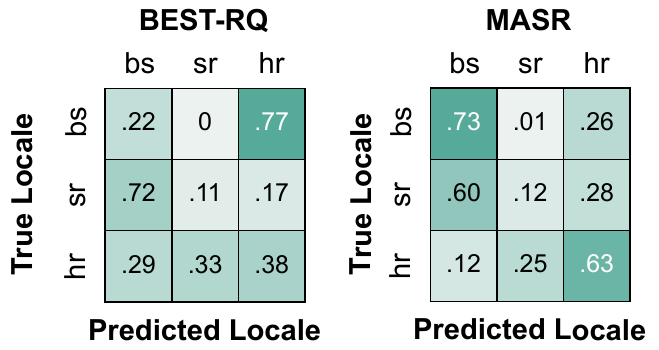}
    \caption{\label{fig:confusion-serbo_croatia} Confusion Matrices for  Serbo-Croatian languages- Bosnian (bs), Serbian (sr) and Croatian (hr). Using syntactic information in pre-training provides gains on these languages. }
\end{figure}

An interesting improvement is that after incorporating the syntactic Lang2vec embeddings, our langID model performs significantly better in disentangling three Serbo-Croatian languages: \textit{Bosnian}, \textit{Serbian} and \textit{Croatian}. As shown in confusion matrix of model predictions on test data (Figure \ref{fig:confusion-serbo_croatia}), the baseline model generates substantial confusions between these languages. However, leveraging the syntactic information of the languages, through the Lang2vec model, we  see an improvement in accuracy for  all languages in this group.

Similarly, substantial improvements were observed in \textit{Japanese} and \textit{
Korean}, where the baseline system predicted $58$\% test utterances of Japanese incorrectly as Korean, while  with MASR pre-training, the error rate dramatically dropped to $3$\%. Aside, $25$\% test examples of \textit{Icelandic} samples had their predictions changed from \textit{
Norwegian}. We also see performance gains in Indic languages such as \textit{Kannada} and \textit{Malayalam}.

The largest performance gains were observed for Southeastern Asian languages, Sub-Saharan African languages, and Southern Asian languages.
For example, $49$\% of \textit{Filipino} utterances were mis-predicted as \textit{Indonesian}, using syntactic distance reduced such mis-classification to $9$\%.
The metadata encoder also helped the prediction of tail languages. It is known that \textit{Galician} and \textit{Asturian} are closely related, as Galician and Asturian are both Romance languages, and spoken in adjacent regions of Spain. 
Syntactic distance helped correct $16$\% of Galician test utterances from incorrect predictions of Asturian, which improved the accuracy by 45\% relative. 

\begin{table}[t]
	\centering
	\small
	\begin{tabular}{lccccccccc}
	    \toprule
	    \multicolumn{1}{c}{\textbf{Method}} & \multicolumn{6}{c}{\textbf{Languages}} & \textbf{Avg} \\
        \cmidrule(r){2-7}
         & de & en & es & fr & it & nl   \\
        \midrule
        BEST-RQ & 5.1 &  7.7 &  5.3 &  6.4 & 10.1 & 11.2 & 7.6  \\
		\ \ + LASR & 4.9 & 7.7 & 5.5 & 6.4 & 10.1 & 10.8 & 7.6\\
        \ \ + \method{} & 5.2 & 7.7 & 5.1 & 6.3 & 10.5 & 11.0 & 7.6\\
		\bottomrule
	\end{tabular}
	\caption{\label{tbl:asr_main} WER (\%) for ASR on Multilingual LibriSpeech using BEST-RQ. Adding   MASR objective during pre-training does not degrade performance on semantic tasks such as ASR.}
\end{table}

\subsection{Locale Level Improvements On Dhwani dataset}
The language identification for Southern Asian languages has been known to be challenging \cite{dataset_fleurs}, the accuracy baselines of langID on Dhwani dataset are thus relative low. The LASR improved the language recognition accuracy by $3$\% on Dhwani. In this work, we proved that, with the inclusion of syntactic similarities, accuracy can be further improved by $2$\%. Our MASR approach made largest improvements in the prediction of \textit{Kashmiri} language, as most predictions that were previously classified as \textit{Urdu} or \textit{Marathi} by LASR system. \textit{Santali} language had a degradation, mainly due to the increased mis-classifications as \textit{Oriya}.

\subsection{Influence of Pre-training Data}

\textbf{Overlapped Languages:}
We have $75$ languages whose speech data are included for pre-training, with $48$ of them being common with languages in the FLEURS dataset, and $5$ being common with those in the Dhwani dataset. We call these languages ``overlapped'' languages. Experimental results show that injecting syntactic similarities improved the language recognition accuracy for most of the overlapped languages, even when they already have sufficient pre-training data. For example, there are about $26$k hours of speech for \textit{German} language, but in the baseline BEST-RQ system, German has a relatively low recall of $21$\% with the major confusion with \textit{Luxembourgish}. The \method{}, with syntactic similarities, successfully corrected most of the false negatives, thus enhanced the accuracy for German by $16$\%. 
Another typical example is Swedish, which has $16$k hours of  speech involved in the pre-training. The baseline system had $45$\% of the \textit{Swedish} test examples predicted incorrectly as Norwegian. MASR reduced such errors to $13$\%, and improved the accuracy of Swedish by $41$\% relative.


More interestingly, even for overlapped languages whose speech pre-training data is limited, the MASR with syntactic similarities enabled to classify them better. For example, the Indian languages whose pre-training speech is less than 50 hours, obtained significant relative improvements in accuracy: \textit{Oriya} (+37\%), \textit{Telugu} (+24\%), \textit{Tamil} (+16\%), \textit{Pashto} (+15\%), \textit{Hindi} (+10\%). Only \textit{Kazakh} and \textit{Georgian} showed performance degradation while the confusions between Kazakh and \textit{Uzbek}, Georgian and \textit{Armenian}, got worsen.

\textbf{Non-overlapped Languages:}
The average relative improvements in accuracy over non-overlapped languages is 14.4\%, larger than the average improvements of 6.7\% on overlapped languages for FLEURS, indicating that MASR framework can effectively leverage the external knowledge base of syntactic similarities, even for languages which do not have any speech data presented in the pre-training, as shown in Figure \ref{fig:delta}. This finding is encouraging, as it may pave the way to scale up language identification to more \emph{unseen} languages, without extra efforts to add significant data resources. 






\section{Discussion and Conclusion}
\subsection{Incorporating Diverse Meta-data}\label{sec:geomasr}
In all the experiments reported thus far, the \method{} setting used only the language label information. As stated in Section~\ref{sec:intro}, one of the goals of formulating the \method{} framework is to facilitate the use of 
multiple streams of label information. 
In line with this goal, we explore the incorporation of additional meta-labels in the form of i) geographic location of the speaker and ii) textual content of the audio data.

For including the meta-data based on the geographic location of the speaker, we added speech recordings from the Vaani dataset\footnote{Vaani Dataset - \url{ https://vaani.iisc.ac.in/}}. This dataset contained audio recordings,  of size $4$k hours belonging to $36$ languages, collected from 80 districts of India. The pre-trained \method{} model with language meta-data was further trained (using the \method{} objective (Equation~\ref{eq:hard-triplet})) with language and geographic location based labels. The meta-data encoder for the language was similar to the one used in previous experiments, while the meta-data for the geographic location used a Haversine distance between the 2-D location vectors containing the latitude and longitude of the speaker's geographic identity. 

For including the meta-data based on textual content, we used the textual transcripts of the audio recordings (when available) in the pre-training dataset. The text content at character level was encoded using an embedding matrix and the average embedding for a given utterance was chosen as the text embedding. Similar to the previous setting using geographic location, the text based \method{} included the encoded text embeddings at the utterance level along with the language labels.  
The NOSS results (Table \ref{tbl:noss}) using   the geographic location based \method{} and the text content based \method{} show that the audio classification is improved using  addition of these meta-labels. However, the speaker and emotion tasks see a degradation as these meta-labels may not be beneficial for those downstream tasks.

\section{Conclusion}
\vspace{-1mm}
In this paper, we have proposed \method{}, a novel approach for incorporating multiple meta-data information sources associated with a speech signal. The objective function of hard mining is modified for the inclusion of these information streams. The final objective function is a combination of the meta-label based objectives with the self-supervision based loss functions. Experiments are performed using speech data along with the meta-data such as language labels, speaker's geographic location and text transcripts. Several downstream evaluations are performed to illustrate the performance improvements achieved using the proposed framework.
\label{sec:summary}

\bibliographystyle{IEEEbib}
\bibliography{strings}

\end{document}